\documentclass[pra,twocolumn,aps,showpacs,10pt]{revtex4}

\usepackage{graphicx}  
\graphicspath{{./Figures/}}
\usepackage{dcolumn}  
\usepackage{amssymb, amsmath}
\usepackage{hyperref}

\begin{document}

\title{Dipolar and scalar $^{3}$He-$^{129}$Xe frequency shifts in stemless cells}
\author{M.\ E.\ Limes}
\author{N.\ Dural}
\author{M.\ V.\ Romalis}
\affiliation{Department of Physics, Princeton University, Princeton, New Jersey, 08544, USA}
\author{E.\ L.\ Foley}
\author{T.\ W.\ Kornack}
\author{A. Nelson}
\author{L.\ R.\ Grisham}
\affiliation{Twinleaf LLC, Princeton, New Jersey, 08544, USA}
\author{J.\ Vaara}
\affiliation{ NMR Research Unit, P.O. Box 3000, FI-90014 University of Oulu, Finland}

\date{\today}
\begin{abstract}

We study nuclear spin frequency shifts in a $^{3}$He-$^{129}$Xe comagnetometer caused by spin polarization of $^{3}$He. We use stemless cylindrical cells to  systematically vary the cell geometry and separately measure the cell shape-dependent and shape-independent frequency shifts. We find that a certain aspect ratio for a  cylindrical cell cancels the dipolar effects of $^3$He magnetization  in the regime of fast spin diffusion.  Using this control we observe  a scalar $^{3}$He-$^{129}$Xe collisional frequency shift characterized by an enhancement factor $\kappa_{\text{HeXe}} = -0.011\pm0.001$ in excellent agreement with theoretical calculation.

\end{abstract}

\maketitle
%

Nuclear spin comagnetometers \cite{aleksandrov1983restriction,Lamoreaux_1986} are used in a number of precision fundamental physics experiments \cite{Safranova_2017}, such as searches for new long-range spin-dependent forces \cite{Bulatowicz_2013,Tullney_2013,Glenday_2008,Venema_1992} and tests of CP, CPT and Lorentz symmetries  \cite{Bear_2000,Rosenberry_2001,XeEDM}. They are also used for  inertial rotation sensing \cite{Donley_2009,Donley_2010,Larsen_2012,Walker_2016,Karlen2018}, and magnetometry \cite{LarsenMag}. 
Measurements of nuclear spin precession frequencies allow nHz level frequency resolution because of long nuclear spin coherence times,  as well as good accuracy and long-term stability \cite{Limes_2017,Sheng_2014}.
 
However, one unavoidable source of frequency shifts in nuclear magnetic resonance experiments is due to magnetic dipolar interactions between  the spins. Local dipolar interactions  are averaged to zero in a gas or liquid due to fast isotropic tumbling, but distant dipolar interactions can lead to complex spin dynamics \cite{Warren_1993,Nacher_1995,Jeener_1999,Ledbetter_2002,Acosta_2008}. Unlike most previous studies, we measure the dipolar   frequency shifts in the regime  where the time scale of atomic diffusion across the whole sample is much faster than  both the time scale of  long-range dipolar interactions and of the transverse spin relaxation.  In this regime, used in most precision co-magnetometer experiments \cite{Sheng_2014}, the frequency shifts depend only  on the shape of the cell containing the atoms. A similar fast-diffusion regime was previously studied by NMR in nanopores \cite{Baugh_2001}.

Nuclear spin dipolar interactions cause systematic frequency shifts in comagnetometer precision measurements  \cite{Allmendinger_2014,Terrano_2018} and are subject of some controversy \cite{Romalis_2014,Allmendinger_2014a}.
Here we use  an anodic bonding batch fabrication process    \cite{Liew_2004}  to make a series of stemless cylindrical cells that contain  $^{3}$He and $^{129}$Xe, as well as $^{87}$Rb and N$_2$. Well-defined cell shapes allow  precision  comparison with a simple theory for dipolar frequency shifts that we develop based on  magnetometric demagnetizing factors \cite{Demag}.  We find that for a certain aspect ratio of the cylindrical cell the dipolar frequency shifts can be eliminated.  An optimal and well-defined cylindrical geometry can improve the stability of nuclear-spin comagnetometers  used for fundamental physics  experiments \cite{Bulatowicz_2013,Glenday_2008,Tullney_2013, Venema_1992,Rosenberry_2001,Allmendinger_2014,XeEDM}. It also can be used in NMR metrology applications instead of spherical cells that are hard to fabricate without stems \cite{Heil_2016}.

Excellent control of long-range dipolar fields also allows us to resolve a small scalar frequency shift between $^3$He and $^{129}$Xe nuclear spins mediated by  second-order electron interactions. Such through-space $J~{\bf I}_1{\bf \cdot}{\bf I}_2$ coupling  in van der Waals molecules has been theoretically studied in several systems \cite{Harris_1998,Pecul_2000,Bagno_2003,Vaara_2013} and was  only observed  experimentally between  $^{129}$Xe and $^1$H in a liquid mixture of Xe and pentane \cite{Ledbetter_2012}. Here we report the first observation of  $J$-coupling between atomic spins in a gas mixture \cite{Limes2018_2}. Simple atomic structure allows   quantitative comparison with a precision first-principles calculation of the  $^3$He-$^{129}$Xe scalar spin coupling strength \cite{Vaara_2018}. We observe a weak temperature dependence of the  scalar enhancement factor $\kappa$, in excellent agreement with calculations. This coupling has been previously assumed to be zero and not considered in precision measurements based on nuclear spin co-magnetometers. 

\begin{figure}
\includegraphics[width=3in]{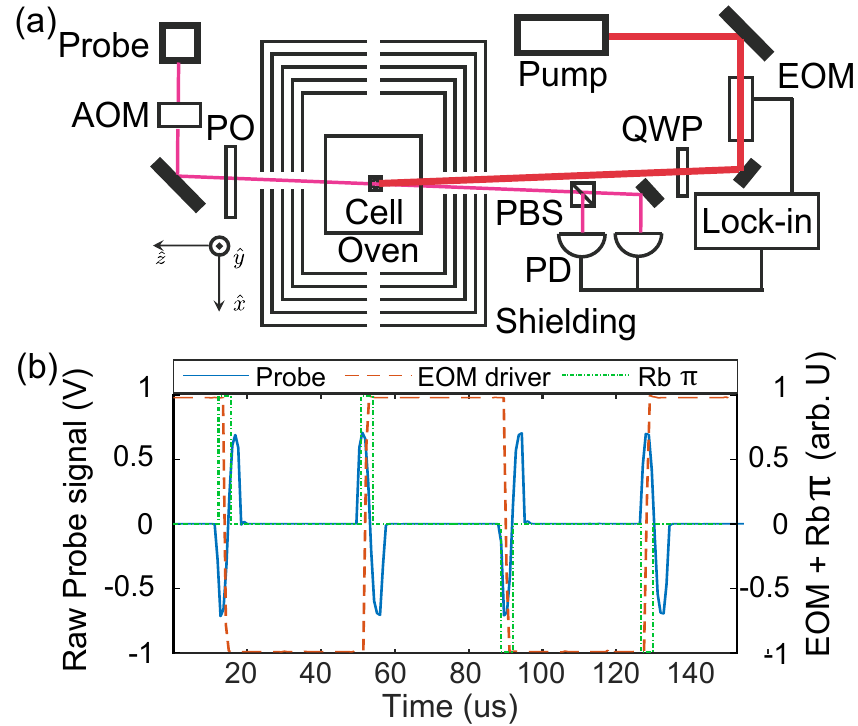}
\caption{(a) Parallel pump-probe pulsed $^{87}$Rb magnetometer with one optical axis along $\hat{z}$. (b) An EOM square wave alternates $\sigma^+$/$\sigma^-$ pump light and $\pi_{\pm y}$ pulses are applied to reverse $^{87}$Rb polarization. The probe laser is gated with an AOM to detect only $^{87}$Rb polarization transitions. A small $B_y$ field changes the $^{87}$Rb transition phase and is detected with a lock-in referenced to half the EOM square wave frequency.}
\label{fig:1}
\end{figure}

 The miniature 6 mm$^3$ vapor cells are made using an anodically bonded glass-Si-glass construction \cite{Liew_2004} in a custom-built system able to fabricate cells containing isotopically enriched alkali metals and noble gases. 
 The Si wafer is 2~mm nominal thickness with a $7\times7$ array of machined  holes with a diameter $d = 2.005\pm0.005$ mm.  We polished one side of a wafer at a small angle to obtain cell height variations from 1.666 mm to 1.988~mm across the wafer.
 The wafer is baked under high vacuum inside the fabrication system to remove contaminants, which is crucial to obtain long wall spin relaxation times,  300 s for $^{129}$Xe and 4 hours for $^3$He. We close the cells with 0.2 mm thick aluminosilicate glass SD-2 that has low $^3$He permeability \cite{Kitchingglass}.  After anodically bonding glass on one side of the wafer, we distill $99.9\%$ isotopically pure $^{87}$Rb metal and bond the second glass in an atmosphere of 80 torr N$_2$, 6.5 torr $^{129}$Xe, and 1400 torr of $^{3}$He. 

The experimental apparatus is shown in Fig.~\ref{fig:1}a. We optically pump $^{87}$Rb  to polarize  nuclear spins by spin-exchange and detect their  precession with an in-situ $^{87}$Rb magnetometer \cite{Sheng_2014, Limes_2017}. This detection yields a high signal-to-noise ratio because Rb experiences enhanced nuclear spin dipolar magnetic field  due  to electron-nuclear Fermi-contact interaction  \cite{Schaefer_1989, Romalis_1998,Ma_2011}. The pulse-train  magnetometer described in \cite{Limes_2017} is adapted here for use with cells with a single optical axis by using counter-propagating pump and probe beams. After an initial pump period to polarize the nuclear spins along a $\hat{z}$ bias magnetic field $B_0 \approx 5$ mG,  the polarization of the on-resonant 795 nm pump laser is switched between $\sigma^+$ and $\sigma^-$ light with an  electro-optic modulator (EOM) at 13 kHz. Simultaneously a train of 3 $\mu$s magnetic field $\pi$ pulses are applied along $\hat{y}$ to flip the $^{87}$Rb polarization back and forth along $\hat{z}$, suppressing  spin-exchange relaxation \cite{Limes_2017}. $^{87}$Rb polarization projection on $\hat{z}$ is detected with paramagnetic Faraday rotation of an off-resonant probe beam passing through the cell to a balanced polarimeter. An acousto-optic modulator turns on the probe laser only during the $\pi$ pulses to avoid unnecessary probe broadening during pumping intervals. An additional $B_y$ field  causes an advancement or retardation of the $^{87}$Rb polarization phase during the $\pi$ pulse flip, see Fig.~\ref{fig:1}b).  The  polarimeter signal is sent to a lock-in amplifier referenced to half the EOM frequency and gives a lock-in output proportional to $B_y$.  The pulse-train magnetometer has a sensitivity of 300 fT/$\sqrt{\text{Hz}}$ in our miniature cells.  

We can also operate the co-magnetometer with $B_0$ perpendicular to the cell's axis. To polarize the nuclear spins we  apply fast $\pi_{\pm y}$ pulses with $\sigma^+$/$\sigma^-$ pumping to create  a time-averaged Rb polarization along $B_0$ parallel to $\hat{x}$. The detection scheme remains the same  because the $\pi$ pulses decouple Rb spin precession due to $B_0$ field \cite{Limes_2017}.

 Each run starts by first polarizing $^3$He for several hours along $B_0$ and then creating a small $ ^{129}$Xe polarization. The pulse-train $^{87}$Rb magnetometer is then turned on and we apply a tipping pulse for the nuclear spins calculated to leave a certain percentage of $M_{\text{He}}$ along $B_0$ and place $^{129}$Xe magnetization in the transverse plane. 
Noble-gas precession signals are recorded in a Ramsey-style sequence for about  100 s, with two detection periods separated by  a dark period that has a rotating, two-axis decoupling pulse train applied \cite{Limes_2017}. 
This decoupling pulse train removes noble-gas frequency shifts due to $^{87}$Rb back-polarization and nulls Bloch-Siegert shifts from the pulse train. 
Each detection period is fit to two decaying sine waves to extract the phases with which the noble gases enter and leave the dark period. 
Knowing the number of cycles elapsed during the dark period, we find the in-the-dark free-precession frequencies of the noble gases $\omega_{\text{He}}$ and $\omega_{\text{Xe}}$.
We divide $\omega_{\text{He}}$ and $\omega_{\text{Xe}}$ by their gyromagnetic ratios $\gamma_{\text{He}}$, $\gamma_{\text{Xe}}$ \cite{Makulski_2015} to find the total effective field  experienced  by $^{3}$He and $^{129}$Xe, $B_0+B_{\text{He}}^d$ and $B_0+B_{\text{Xe}}^d$, respectively.

\begin{figure}
\includegraphics[width=3in]{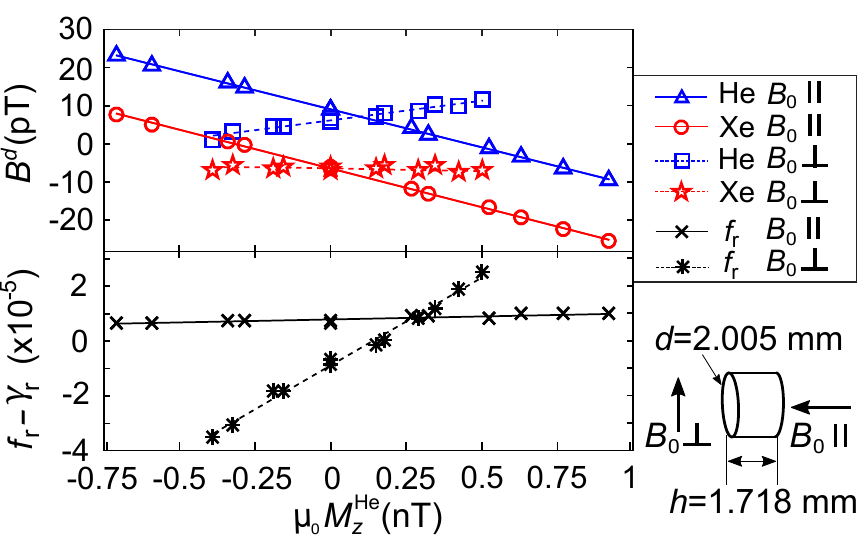}
\caption{Top panel: the dipolar field $B^{d}$ experienced  by $^{3}$He (triangles) and $^{129}$Xe (circles) from $^{3}$He magnetization $M_{\text{He}}$ along $B_{0}$  $\parallel$ to the cylinder axis and for  $^{3}$He (squares) and $^{129}$Xe (stars) for $B_{0}\perp $ to the cylinder axis with $h/d = 0.857$. Lines show linear fits. Bottom panel: The comagnetometer frequency ratio $f_r -\gamma_r = \omega_{\text{He}} /\omega_{\text{Xe}}-\gamma_{\text{He}}/\gamma_{\text{Xe}}$ is shown for $B_{0} \parallel$ ($\times$'s) and $B_0 \perp$ ($*$'s) to the cylinder axis.}
\label{fig:2}
\end{figure}

After a Ramsey measurement we place $M_{\text{He}}$ along or against $B_0$ by using dumping feedback \cite{Alem_2013}, and repeat the sequence for many $M_{\text{He}}$ values and tipping angles to find the slopes $B^d/\mu_0 M^{\text{He}}_z$ for $^{3}$He and $^{129}$Xe.
To avoid systematic errors from $B_0$ drift, chemical shifts, and  remnant $^{129}$Xe polarization projection onto $B_0$ we use identical tipping pulses while alternating the sign of the initial $M_{\text{He}}$ projection along $B_0$. 
$M_{\text{He}}$ is found by comparing the amplitude of the fitted $^{3}$He signal  to the  $^{87}$Rb magnetometer response from a known external magnetic field.
The $^3$He signal measured by  Rb atoms  is given by $B_{\rm Rb}=2 \mu_0 \kappa_{\rm RbHe}M_{\rm He}/3$  \cite{Romalis_1998}.  We neglect a 1\% correction from the long-range dipolar effect of $M_{\text{He}}$ on $^{87}$Rb. In Fig.~\ref{fig:2} we  plot the effective dipolar field experienced by $^{3}$He and $^{129}$Xe due to  $M_z^\text{He}$  for a cell with $h=1.718$ mm, $d=2.005$ mm ($h/d = 0.857$) at 120$^{\circ}$C. We make the measurements with $B_0$  parallel and perpendicular to the optical axis of the cylindrical stemless cell.

In our cells  the diffusion time across the cell is much shorter than the period of  spin precession in dipolar fields and  the spin relaxation time. In this regime each spin species has a uniform nuclear magnetization $\bf M$ inside the cell, unlike previous studies of long-range dipolar fields \cite{Grover_1990,Warren_1993}. The spin precession frequencies are determined by the volume average magnetic field inside the cell, which can be calculated using the magnetometric demagnetizing factors for the average $\bf H$ field in a uniformly-magnetized sample, $\left \langle H_i\right\rangle_V=-n_i M_i$, $(i=x,y,z$) \cite{Demag}. Analytical expressions for $n_i(\gamma)$ for a cylinder, where $\gamma=h/d$ is the height over diameter ratio, are given in \cite{Joseph_1966,Chen_2006}. The demagnetizing factors satisfy $n_x+n_y+n_z=1$. For a cylinder with $\gamma=0.9065,$ $n_i=1/3$ (the same as for a sphere)---for this aspect ratio the dipolar effect is zeroed out for all orientations of $B_0$ with respect to the cylindrical axis.

The classical average magnetic field needs to be corrected by separating out the scalar contact term $2 \mu_0\delta^3({\bf r}){\bf m}/3$   of the classical dipolar  field for non-interacting point dipoles $\bf m$ \cite{Jeener_1996}. For real particles, the contact interaction can be enhanced or suppressed relative to the classical value \cite{Romalis_2014}.  It is enhanced for electron-nuclear spin interactions between alkali metal and noble gas spins, $\kappa_{\rm RbHe}=5.6, \kappa_{\rm RbXe}=500$  \cite{Schaefer_1989,Romalis_1998,Ma_2011}. For nuclear spins the direct contact interaction is zero,  but indirect second-order electron-mediated $J$-coupling can induce  a finite effect.  The scalar spin-spin interaction can be generally parametrized  by $\kappa$, where $\kappa=1$ would correspond to the classical result \cite{Schaefer_1989,Heckman_2003}. We write  
\begin{equation}
 \left\langle B^{d}_i\right\rangle_V=\mu_0\left[ M_i-n_i M_i+\frac{2}{3}(\kappa-1)  M_i \right ].
\end{equation}

Bloch equations can be used to describe spin precession in the presence of  a constant bias field and small nuclear-spin dipolar fields. In our case only $^3$He has a significant magnetization. The dipole field experienced by $^{129}$Xe
from $^3$He can be written as $\left\langle B^{\text{Xe}}_z\right\rangle_V=\mu_0 (1/3-n_z +2\kappa_\text{HeXe}/3)M_z^{\text{He}}$, where the $z$ axis is defined by the magnetic field direction. The rotating components of the $^3$He magnetization do not have a net effect on $^{129}$Xe precession frequency to first order in $M_z^{\text{He}}$.  From the linearity of data in Fig.~2 one can see that the second-order Bloch-Siegert shift from the rotating components is negligible in our case.

In contrast, the $^3$He precession frequency is affected by the secular co-rotating components of the $^3$He dipolar field but is not affected by the scalar contact interaction with $^3$He because scalar interactions are not observable between like spins. One can write \cite{Jeener_1996,Warren_1995}:
\begin{equation}
\frac{\left\langle {\bf B}^{\text{He}}\right\rangle_V}{\mu_0}= 
\left(\frac{n_z}{2}-\frac{1}{6}\right){\bf 
M}^{\text{He}}+\frac{3}{2}\left(\frac{1}{3}-n_z\right) M_z^{\text{He}}\hat{z}.
\end{equation}
The first term on the right hand side does not generate any frequency shift because it gives $\left\langle {\bf B}^{\text{He}}\right\rangle_V$  parallel to ${\bf M}^{\text{He}}$.  The effective dipolar field responsible for a $^{3}$He  frequency shift  is  given by the second term. It is  3/2 times larger than for $^{129}$Xe and both are proportional to the $M_z^{\text{He}}$ projection.

\begin{figure}
\includegraphics[width=3in]{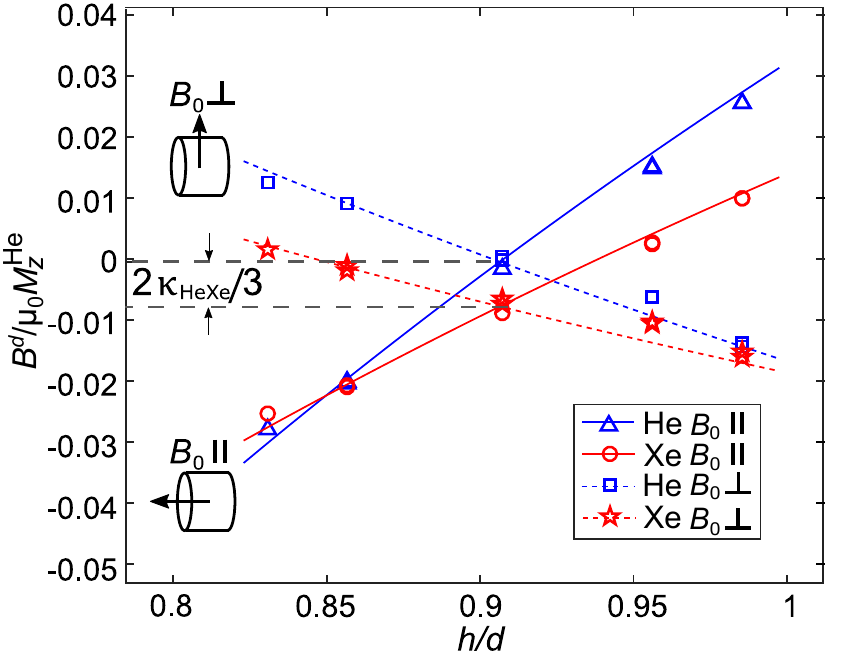}
\caption{The slope $B^{d}/\mu_{0}M^{\text{He}}_z$ is plotted against the cell aspect ratio $h/d$ for $B_0 \parallel$ and $B_0 \perp$ to the cylinder axis. Lines show dipolar field theory with only $\kappa_{\text{HeXe}}$ as a free parameter.
}
\label{fig:3}
\end{figure}

\begin{table}
\begin{tabular}{|c|c|c|c|c|}
\hline
~ &$^{129}$Xe field &$^{129}$Xe theory &$^{3}$He field&$^{3}$He theory\\
\hline
$B_0 \parallel$ & $1 $ &$1$  &$1.50\pm0.02$  &$3/2$\\
$B_0 \perp$ & $-0.44 \pm 0.03$  &$-1/2$&  $-0.72 \pm 0.03$&  $-3/4 $\\
\hline
\end{tabular}
\caption{Relative size of the slopes $(B^d/\mu_0M^{\text{He}}_z)/(h/d)$ from fits to Fig.~\ref{fig:3} data, scaled to the $^{129}$Xe $B_0 \parallel$ case. }
\label{t:table}
\end{table}

In Fig.~\ref{fig:3} we plot the slope of the dipolar field  $B^d/\mu_0 M^{\text{He}}_{z}$ for a number of cells as a function of the cell aspect ratio $h/d$, {\em e.g.}~the slopes of the line fits in the top panel of Fig.~\ref{fig:2} are shown at  $h/d = 0.857$. 
 The lines through the $^{3}$He data in Fig.~\ref{fig:3} represent first-principles calculations of the dipolar fields using the theoretical expression for $n_z(\gamma)$ for a cylinder. For Xe data the lines $B^d_\text{Xe}/\mu_0 M^{\text{He}}$ have a vertical offset  representing the scalar  frequency shift $\kappa_\text{HeXe} = -0.011 \pm 0.001$, where the error is determined by the $M_z^\text{He}$  calibration 
uncertainty.
 
The accuracy of the dipolar field model in  cells with well-defined geometry can also be checked by fitting the data points in Fig.~\ref{fig:3}. The relative size of the long-range dipolar frequency shifts for the four cases are related to each other by simple ratios shown in Table~\ref{t:table}, which are applicable for any cell with uniaxial symmetry \cite{Demag}.  The  $^{3}$He  data cross zero and the  $^{129}$Xe data intersect each other at $h/d=0.911\pm0.001$, while $n_z(\gamma)=1/3$ for a cylinder with $h/d=0.9065.$ The $0.5\%$ discrepancy is likely due to a slightly larger effective diameter of the holes drilled in the Si wafer, which have a surface roughness of about $10~\mu$m and were measured with pin gauges that determine the minimum diameter.

The existence of a finite $\kappa_\text{HeXe}$ implies that the cell aspect ratio required to cancel   the nuclear  magnetization effects on the frequency ratio in the comagnetometer  is different  from  the condition $n_i=1/3$.  The optimal cell aspect ratio depends on the orientation of $B_0$ with respect to the cylindrical axis.  In the bottom panel of Fig.~\ref{fig:2} we show the comagnetometer frequency ratio $f_r-\gamma_r\equiv\omega_\text{He}/\omega_\text{Xe}-\gamma_\text{He}/\gamma_\text{Xe}$ for the cell with $h/d = 0.857$. We find $f_r$ is almost independent of  $\mu_0 M^{\text{He}}$ for $B_0$ parallel to the cylinder axis for this particular cell.

\begin{figure}
\includegraphics[width=3in]{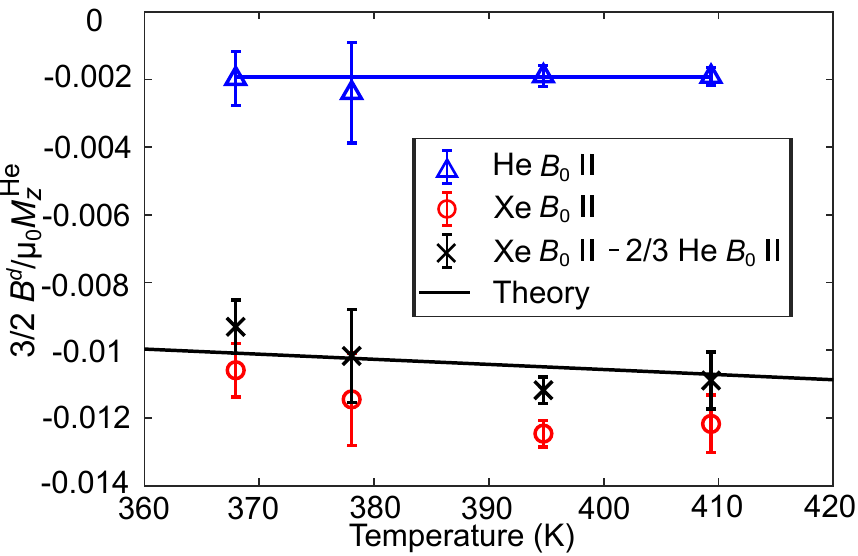}
\caption{Temperature dependence the dipolar and scalar effects for a cell with $h/d = 0.905$, a geometry close to nulling the dipolar fields. A line through the $^{3}$He data (triangles) shows the remnant dipolar field due to $M^{\text{He}}_z$ is independent of temperature. The remnant dipolar field is subtracted from the $^{129}$Xe data (circles) to find the temperature dependence of the scalar $\kappa_{\text{HeXe}}$ interaction.  First-principles calculations for $\kappa_{\text{HeXe}}$ are shown with black line. }
\label{fig:4}
\end{figure}

We measured  the temperature dependence of $\kappa_\text{HeXe}$ using  a cell with $h/d = 0.905$. In Fig.~\ref{fig:4} we plot $3B^d_\text{}/2\mu_0 M^{\text{He}}$ which is equal to  $\kappa_\text{HeXe}$ for $^{129}$Xe in the absence of dipolar field. As  expected, the small remaining dipolar field for $^{3}$He  is independent of temperature. We use the $^{3}$He data and the relative slope from Table~\ref{t:table} to correct the $^{129}$Xe data for the remaining dipolar field. Relatively weak temperature dependence of $\kappa_\text{HeXe}$  implies that dipolar and scalar frequency shifts can cancel each other  over a significant temperature range
by proper choice of the cell shape and orientation.

The indirect spin-spin $J$ couplings for noble gases was calculated recently from first-principles for $^{3}$He-$^{129}$Xe \cite{Vaara_2018}, and previously for $^{129}$Xe-$^{131}$Xe \cite{Vaara_2013}.
  For freely moving spins in gases and liquids  it is more convenient to parameterize  the interaction in terms of $\kappa$ \cite{Heckman_2003,Ledbetter_2012}:
\begin{equation}
\kappa=-\frac{3 \pi}{\mu_0 \gamma_1 \gamma_2 \hbar} \int 4 \pi r^2 g(r)J(r)dr
\end{equation}
 where $g(r)$ is the radial intermolecular distribution function. 
Ref.~\cite{Vaara_2018} gives $\kappa_{\text{HeXe}}=-0.0105\pm0.0015$ at a temperature of 120 $^{\circ}$C, in excellent agreement with our experimental value. 
A temperature dependence is also predicted as $\kappa_{\text{HeXe}} = -0.00453-1.51\times 10^{-5}T$ with $T$ in K. While small, this temperature trend is in good agreement with  our data.

In conclusion, we have used stemless cells with well-controlled geometry to investigate dipolar and scalar frequency shifts in nuclear spin co-magnetometers. We developed a simple model for dipolar magnetic fields  based on magnetometric demagnetizing factors that applies in the regime of fast diffusions. It is in excellent agreement with our measurements. We observed  a small scalar interaction between $ ^{3}$He and $^{129}$Xe spins and find it to be in very good agreement with recent first-principles calculation~\cite{Vaara_2018}. The presence of  scalar spin-spin coupling implies that the optimal shape of a co-magnetometer cell is slightly different from a sphere. We show that a cylindrical cell with a certain aspect ratio and orientation relative to the bias field can cancel both dipolar and scalar frequency shifts. Thus, one can  eliminate an intrinsic source of frequency shifts present  in all nuclear spin co-magnetometer experiments.

This work was funded by DARPA, NSF, and  Academy of Finland.

\end{document}